\documentclass{emulateapj} %apt-get update; apt-get install emulateapj

\slugcomment {Accepted by the Astrophysical Journal Letters}
\shorttitle{New probe of inner T Tauri disks}
\shortauthors{Bergin et al.}

\newcommand\msunyr{\rm M_{\odot}\,yr^{-1}}
\newcommand\msun{\rm M_{\odot}}

\newcommand\mdot{ \dot{M}}

\newcommand\be {\begin{equation}}
\newcommand\en{\end{equation}}
\newcommand\cm{\rm cm}

\newcommand\dm{\rm DM~Tau}

\newcommand\tw{\rm TW~Hya}
\newcommand\gm{\rm GM~Aur}
\newcommand\lk{\rm LkCa~15}

\def\'#1{\ifx#1i{\accent"13\i}\else{\accent"13#1}\fi}

\def\10alamenos#1{10$^{-#1}$}
\def\10ala#1{10$^{#1}$}

\def\cm2g{\rm cm^2 \ g^{-1}}

\begin{document}

\title
{
A New Probe of the Planet-Forming Region in T Tauri Disks
}

\author{Edwin Bergin\altaffilmark{1} and Nuria Calvet\altaffilmark{2}, 
Michael L. Sitko\altaffilmark{3,10}, 
Herve Abgrall\altaffilmark{4},
Paola D'Alessio \altaffilmark{5},
Gregory J. Herczeg\altaffilmark{6},
Evelyne Roueff\altaffilmark{4},
Chunhua Qi\altaffilmark{2},
David K. Lynch\altaffilmark{7,10}, Ray W. Russell\altaffilmark{7,10},
Suellen M. Brafford\altaffilmark{8,10}, R. Brad Perry\altaffilmark{9,10}}
%Paola D'Alessio \altaffilmark{2}, and Gregory J. Herczeg\altaffilmark{3}}

\altaffiltext{1}{University of Michigan, 825 Dennison Building, 501 E. University Ave.,
Ann Arbor, MI 48109-1090; email: ebergin@umich.edu}
\altaffiltext{2}{Harvard-Smithsonian Center for Astrophysics, 60 Garden St.,
Cambridge, MA 02138}
\altaffiltext{3}{Department of Physics, University of Cincinnati, Cincinnati OH 45221-0011}
\altaffiltext{4}{LUTH, UMR 8102 du CNRS,
Observatoire de Paris, Section de Meudon, 
Place Jules Janssen, 92195 Meudon, France}
\altaffiltext{5}{Instituto de Astronom\'{\i}a,
UNAM, Apartado Postal 72-3 (Xangari), 58089 Morelia, Michoacan, Mexico;
M\'exico}
\altaffiltext{6}{
JILA, University of Colorado, Boulder, CO 80309-0440}
\altaffiltext{7}{The Aerospace Corporation, Los Angeles, CA 90009}
\altaffiltext{8}{School of Law, University of Dayton, Dayton, OH 45469-2760}
\altaffiltext{9}{NASA Langley Research Center}
\altaffiltext{10}{Visiting Astronomer, NASA Infrared Telescope Facility, 
operated by the University of Hawaii under contract with the National 
Aeronautics and Space Administration.}

\begin{abstract}
We present new observations of the FUV (1100-2200 \AA ) radiation field
and the near- to mid-IR (3--13.5 $\mu$m) spectral energy
distribution (SED) of a sample of T Tauri stars selected on the basis of bright
molecular disks (GM Aur, DM Tau, LkCa15).   In each source we find
evidence for Ly $\alpha$ induced H$_2$ fluorescence and an additional
source of FUV continuum emission below 1700 \AA .  Comparison of the
FUV spectra to a model of H$_2$ excitation suggests that 
the strong continuum emission is due to electron impact excitation of H$_2$. 
The ultimate source of this excitation is likely X-ray irradiation which
creates hot photo-electrons mixed in the molecular layer.
Analysis of the SED of each object finds 
the presence of inner disk gaps with sizes of a few AU in 
each of these young ($\sim$1 Myr) stellar systems.   
We propose that the presence of strong H$_2$ continuum
emission and inner disk clearing are related by the increased penetration
power of high energy photons in gas rich regions with low grain opacity. 
%Thus the UV emission from H$_2$ is a probe of the gas conditions within
%the planet forming regions of T Tauri disks. 
\end{abstract}


\section{Introduction}
\label{sec_intro}
\setcounter{footnote}{0}

In recent years there has been growing evidence for evolution of solid
particles in young 
%NC
($\le 10$ Myr) proto-planetary accretion disks 
\citep[][and references therein]{beckwith_ppiv, dalessio03}.   
The onset of this evolution lies in the coagulation of sub-micron
sized particles into larger grains; a process which continues
until the larger grains decouple from the gas and 
settle to a dusty mid-plane.  In the mid-plane these solid particles
grow in size until they become large enough to gravitationally
focus collisions with smaller bodies, eventually making planets
(Safronov 1972; Weidenschilling 1997). 

Since dust grains within the disk absorb stellar UV and optical photons
and re-emit at wavelengths longer than $\gtrsim 2 \mu$m, the
dust evolutionary process has direct consequences on disk continuum emission.
Thus, at early evolutionary stages the presence of dust in the inner disk 
($\lesssim 10$ AU) is revealed by 
optically thick emission at near and mid-infrared (IR) wavelengths. 
%and the outer disk
%by optically thin emission at millimeter and sub-millimeter wavelengths.
%This emission represents the so-called excess above stellar photospheric emission.   
As grains grow
the opacity at these wavelengths decreases, 
revealing stellar photospheric emission.
%Thus the emission excess and its wavelength dependence has 
%been widely used to characterize
%disk populations and lifetimes \citep{skrutskie_10mic, haisch_dlife,
%meyer_iso}, temperature 
%and mass \citep{beckwith_mass, dutrey_mass}, and planet formation/grain growth 
%\citep{beckwith_beta, dalessio_disk, chiang_disk, calvet_twhya}.
The formation of giant planets
can also affect the dust emission.
%\citep{lin_pap86, lin_pap93,
%bryden99}.  
Gravitational interaction between the disk
and the forming planet results in the formation
of a gap as the mass of the planet increases
(Bryden et al. 1999; Alibert, Mordasini \& Benz 2004 and references therein),
%(Nelson \& Papaloizou 2003;
%Rice et al. 2003; Bate et al. 2003; Winters et al. 2003; Alibert et al. 2004)
producing a significant decrease in disk flux in the near-IR
(Rice et al. 2003, R03).
%These connections between dust evolution in the disk and its observational
%signatures, will become more evident 
%with the anticipated results from the Spitzer Space Telescope. 

What is less recognized, at least in terms of an observational signature, 
is that the
molecular evolution is closely linked 
with the dust evolution.
Disks in a more advanced degree of dust evolution
will be more easily permeated
by the destructive short-wavelength radiation fields generated
in part by accretion.    Since grain evolution and
planet formation proceeds more rapidly in the denser 
inner disk (Weidenschilling 1997), these effects will be magnified in the very regions that are
closest to the source(s) of radiation. 
In this {\em Letter} we suggest that a strong H$_2$ UV emission feature 
is an observational consequence  
of grain growth/planet formation in the inner disks of young
($\sim 10^6$ yr) accreting T Tauri disks.  
%Thus H$_2$ UV emission is a probe of the conditions within the
%planet-forming regions of proto-planetary disks.

\section{Observations}
%\subsection{HST/STIS UV Spectra}

The sources chosen for the STIS UV study are \dm , \gm , and \lk . 
In Table~1 we provide some basic characteristics for these systems.
All sources have relatively similar properties and were selected on the
basis of the presence of 
%NC
gas disks with
rich
molecular complexity \citep{koerner_gmaur,
dutrey_diskchem, qi_lkca15}.  
Each are single star systems with no evidence for binary companions
\citep{white_binary}.
HST/STIS spectra of \dm , \lk , and \gm\ were obtained for HST program G09374
on Feb. 5, Feb. 13, and Apr. 1, 2003 respectively. 
Exposures were taken using the G140L (1150 \AA\ $-$ 1730 \AA ) and G230L
(1570 \AA\ $-$ 3180 \AA ) gratings with an aperture size of 2$''$. The
spectral resolution per pixel for G140L is $\Delta \lambda =$ 0.6 \AA\ and
$\Delta \lambda =$ 1.58 \AA\ for G230L. With a FWHM of the PSF of 1.5 pix
at the FUV and 2 pix at the NUV, this results in effective resolutions of
 0.9 \AA\ (R$\sim$1550) at the FUV and 3 \AA\ in the NUV (R$\sim$ 770).
The plate scale is 0.024$''$ per pixel in both FUV and NUV, so the aperture
size is large enough for spectrophotometry.
Exposure times were 10m (G230L) and $\sim$65m (G140L) for \dm\ and \gm\ and
189m (G140L) and 37m (G230L) for  
\lk .
Standard CALSTIS pipeline procedures were used to reduce the 
data.\footnote{There is some overlap 
in wavelength coverage between the NUV and FUV detectors.
No correction was needed to be applied to match flux levels, and the data is
presented with overlap included.  Some data was not shown 
at the short wavelength end of the NUV spectra to better illustrate particular features.
However, the NUV data in this spectral region closely mirrors the FUV data. }

%\subsection{Mid-Infrared Spectrophotometry}

%NC
The targets were observed in the mid-IR over the
%The mid-IR observations of these T Tauri stars were observed over the
course of three separate observing runs (Jan. \& Feb. 2003) that overlapped the
HST observations for \dm\ and \lk\ (5-6 week difference for \gm ).
Observations were obtained with the Aerospace
Corporation's Broadband Array Spectrograph System (BASS; Sitko, Lynch, \& Russell 2000) with a 3.4$''$ beam
on the NASA Infrared Telescope Facility.
%using the NASA Infrared Telescope Facility (IRTF)
%were obtained using the Aerospace 
%Corporation's Broadband Array Spectrograph System (BASS) with a 3.4$''$ beam. 
%NC i replaced a reference for the description 
%Description of the instrument is given in \citet{sitko00}.
This instrument 
uses a cold beam-splitter to separate the light into two separate wavelength 
regimes (2.9-6 $\mu$m; 6-13.5 $\mu$m). Each beam is dispersed onto a
58-element Blocked Impurity Band linear array, thus allowing for 
simultaneous coverage from 2.9-13.5 $\mu$m. The spectral 
resolution is wavelength-dependent, ranging from 
R $\sim$ 30 to 125 over each wavelength region.
Integration times were 40 min (GM~Aur), 227 min (DM Tau), and 143 min (LkCa15).
All observations are calibrated relative to $\alpha$ Tau, with 
typical airmasses $\sim 1.07$ and calibrator airmasses ranging from 1.03--1.12.
%\footnote{For objects where multiple observations were obtained (LkCa 15 and 
%DM Tau), the data were checked for possible variability. No significant 
%variations were detected in LkCa 15. The data on DM Tau (at 0.03 Jy) were 
%of insufficient quality to make this assessment. In both cases, the data 
%were combined merged to provide the cleanest spectra possible.}

\section{Results}

\subsection{UV Spectra of T Tauri Stars}

Figure~1 shows the FUV de-reddened fluxes of the T Tauri stars in our sample
including the spectra
of \tw\ from \citet{herczeg_twhya1}. 
%and BP Tau \citep{bergin_lyalpha}.
%Of particular importance is the strength of the overall  FUV radiation field.
%These stars are characterized by strong, and often chemically diverse,
%molecular emission that is believed to be 
%controlled by energetic ultraviolet (UV) and X-ray
%radiation  \citep{vanz_zchem}.
In Table~1 we provide the strength of the FUV radiation at 100 AU (estimated
by integrating the FUV radiation field from 1100 to 1700 \AA ) 
normalized to that of the interstellar UV radiation field.
This estimate is a lower limit due to the unknown 
Ly~$\alpha$ flux.\footnote{There 
is little information of the strength of the radiation
field below 1100 \AA .   
%NC
FUSE data of TW Hya indicates that the
%Some information exists from FUSE data on TW Hya that the
FUV field strength, G$_0 =$ 3400 (including Ly $\alpha$), 
does not change appreciably with this radiation included
\citep{herczeg_twhya2}.  
%Moreover, the estimates are certainly good in a relative sense between
%each object. 
}

\begin{figure}
\epsscale{1.3}
\plotone{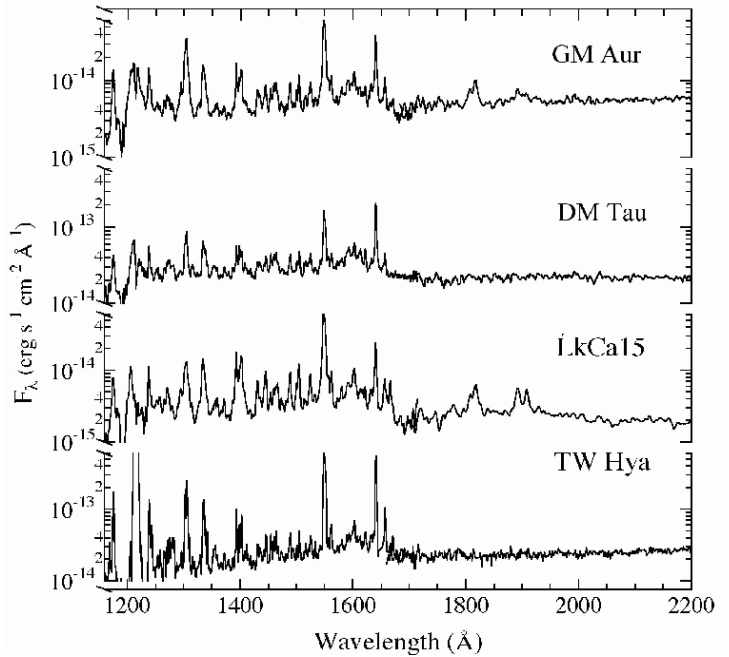}
\caption{HST/STIS spectra of objects in our sample, including
the spectra of TW Hya first discussed in \citet{herczeg_twhya1}, but
with reduced spectral resolution (1.5 \AA ).
%NC and BP~Tau.
}
\end{figure}

%In the following we will not discuss atomic line emission features beyond HI \lya .  
Figure~2 shows the spectrum of LkCa15 with prominent emission features
identified.
In \tw\ (Figure~1) there is a strong Ly $\alpha$ emission line that
is not as evident in the other sources,
because these sources are embedded within molecular clouds; in contrast 
\tw\ has no local cloud and little interstellar absorption \citep{herczeg_twhya2}. 
Evidence for strong Ly $\alpha$ radiation that penetrates to the molecular layer
can be inferred from the presence of numerous
emission features coincident with Ly $\alpha$ pumped H$_2$ emission lines.  
In Figure~2 some of these
coincidences are denoted for \lk, but similar features are seen in all sources shown
in our sample.   
Beyond the clear H$_2$ emission line features there also appears a sharp rise
in emission below 1700 \AA .
This feature 
is likely due to a combination of {\em discrete and continuum}
emission emitted by H$_2$ molecules in excited electronic states.
%The inference of strong continuum emission is not due to 
%the lack of proper spectral resolution as the H$_2$ continuum is
%clearly seen in the higher resolution TW Hya spectrum 
%which was obtained by STIS using the E140M grating
%Moreover the continuum rise is coincident
%wit the wavelength
%regime with overlapping coverage in the STIS G140L and G230L gratings and
%is the feature is seen in both independent measurements.

H$_2$ emission below 1700~\AA\ can result from at least
two physical mechanisms (cf. Herczeg et al. 2004), Ly~$\alpha$ induced fluorescence and 
electron impact excitation followed by fluorescence (Liu et al. 2002).  
Ly $\alpha$ excitation involves only the 
ungerade electronic states (B $^1\Sigma_u$ and C $^1\Pi_u$),
producing  a spectrum permeated by discrete emission lines
with some continuum emission.
However, electron impact excitation can also excite 
the gerade states (E, F etc.) which cascade toward  
the B, C states and emit subsequent UV fluorescence. 
This produces a broader spectrum 
with greater contribution from the H$_2$ dissociation continuum  
(Abgrall et al. 1997; Jonin et al. 2000; Liu et al. 2003).
%Such effects have been studied in Liu et al. (2002).
%Both model spectra show that the rise in emission begins near $\sim$1700~\AA , with
%a peak in the continuum near 1600 \AA,
%as in the HST data for \gm , \dm , \lk\ and \tw . 
A model spectrum for 100eV electron impact spectrum
is displayed in Figure 2.
%\tw\ is a factor of $\sim 2$ closer than the Taurus-Auriga systems and suffers
%from little extinction as it is not associated with a molecular cloud. 
%These factors allowed 
%\citet{herczeg_twhya1, herczeg_twhya2} to examine the $H_2$ UV emission  
%from \tw\ in great detail.  They concluded that the emission is primarily the
%result of Ly $\alpha$ induced fluorescence.  However the increase in continuum flux 
%below 1700 \AA\ cannot be explain by \lya\ fluorescence.  Instead it  
%is attributed to either electron collisions, H$_2$ fluorescence by FUV continuum emission,
%or a hot accretion component.   
A simple comparison of our data to the 
electron impact model in Figure~2 shows similarities that are
suggestive  that electron impact excitation may be important.\footnote{In 
the electron impact model there is an additional strong feature near 1200 \AA\ which
we do not discuss due to the close proximity of Ly $\alpha$.  In a future publication  we will compare the FUV spectra
with H$_2$ excitation models in greater detail.}
%\footnote{Comparison 
%with the high resolution \tw\ spectrum with
%the contribution from \lya\ fluorescence subtracted using the results of
%\citet{herczeg_twhya1}, reveal the presence of residual line emission at  1608 \AA.
%This line coincides with the strongest line from the electron impact model,
%providing additional evidence for electron excited H$_2$ UV emission.}
%We are in the process of a deeper analysis of these data using
%an extensive grid of models, which will
%compare the relative importance of electron impact excitation and 
%\lya\ fluorescence on the observed UV spectra of stars in our sample.  This
%will be provided in a subsequent publication.  
%The inference of 
%H$_2$ electron impact excitation has impact beyond the physics of
%H$_2$ collisions.  It 
%potentially provides the first link between X-ray emission,
%producing hot photo-electrons in the disk,
%and a significant portion of the FUV radiation field
%during a stage where planets may just
%be forming within the disk.   
The implications of this result are discussed in \S 4.

\begin{figure}
\epsscale{1.1}
\plotone{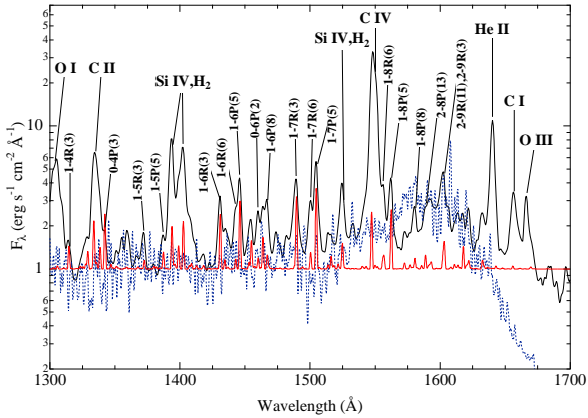}
\caption{
Comparison of  observed LkCa15 spectrum (solid black line), with strong features identified,
the electron impact model spectrum (dotted blue line), and model of Ly $\alpha$ fluorescence
(solid red line).  
The model spectra were produced
using an excitation model and molecular transition probabilities 
in the discrete and continuum range (Abgrall, Roueff, \& Drira 1994, 2000). 
For ease in comparison, the
fluxes from both observation
and models are normalized to the flux at 1425 \AA .
}
\end{figure}

\begin{figure*}
\epsscale{1.1}
\plotone{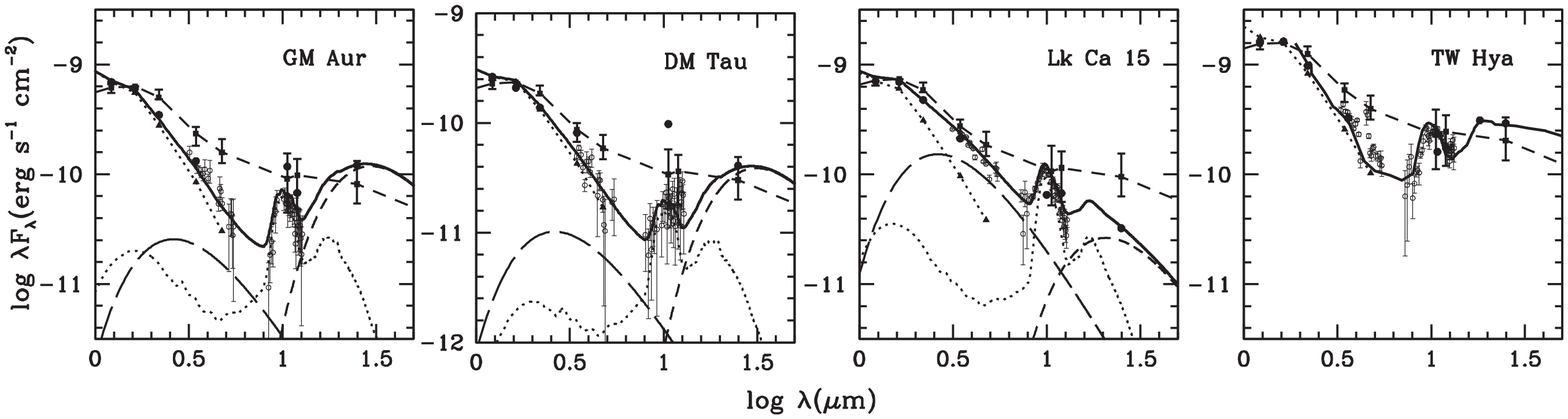}
%\plotone{/home/ncalvet/md2/paola/tw/modmay04/fig_ted_comp.ps}
\caption{Spectral energy distributions of \gm, \dm, and
\lk: BASS data (small open circles),
optical and IRAS bands from KH95 (large solid circles).
Also shown are the
photospheric fluxes (solid triangles joint by dotted line),
and the contribution of the wall (short-dashed lines),
the optically thin inner region (dotted line), and
the possible rim at the dust destruction radius
(long dashed line). We show for comparison
the median of Taurus (solid rectangles joined by
dashed lines, error bars denote the
quartiles of the distribution) from D'Alessio et al. (1999).
The last
panel shows the SED of TW Hya, with data and
model from Calvet et al. (2002).
Photospheric fluxes are calculated from standard colors and spectral
types from KH95, scaled to the de-reddened J magnitude
of each object.
Models discussed in \S 3.2 are shown in solid lines.
We do not include the contribution of the
outer disk, which becomes important at longer
wavelengths that those considered here (C02).
}
\end{figure*}
\subsection{Spectral Energy Distributions}

The targets, selected on the basis of strong disk
molecular emission, show indications of significant
dust evolution in their inner disks. For example,
the $K-L$ colors of {\lk} and {\gm} are bluer than 86\% of the
Taurus sample (consisting of
51 CTTS with near and mid-IR observations from 
Kenyon \& Hartmann 1995, KH95),
and the $K-L$ color of {\dm} is bluer than 70\% of the sample.
The BASS observations of the targets
strengthen these indications.
In Figure 3, we compare the mid-IR, optical, and IRAS data
of the targets to the 
median SED
of Classical T Tauri stars in
Taurus from D'Alessio et al. (1999). The
SEDs of the targets show
a clear flux deficit in the near-IR
relative to the median in Taurus, analogous
to that of TW Hya (cf. Figure 3), for which
inner disk clearing
due to the action of a planet opening
a gap in the disk has been suggested (\S 1; C02; R03).

Attempting to get insight into the 
structure of the inner region, we use a simple
model, following
C02 and Uchida et al. (2004). The model
consists of:
(1) A vertical ``wall'' in the inner edge
of the outer disk, representing
the far edge of the gap. This wall, located at
radius $R_t$ with height $z_w$ is
frontally illuminated by the star and emits
as a black body with temperature $T_w$,
where $T_w = T_{eff}/R_t^{1/2}/2^{1/4}$ (cf. C02).
We are ignoring effects of geometry
of the gap, inclination, and occultation.\footnote{Emission from the
wall in the disk is inclination dependent. However,
the targets have inclinations $\sim$ 30 - 60$^\circ$ \citep{simon00,qi_lkca15},
for which the wall emission is within 50\% of the maximum \citep{dull01}.}
(2) An optically thin inner disk, extending
%from $R_{min} \sim R_d \sim$ 0.07 AU, the dust
from the dust
destruction radius to $R_t$, with optical
depth $\tau_{min}$ at 10 $\mu$m. This region
has a mixture of small ($\sim 0.1 \mu$m) and 
large ($\sim 2 \mu$m) grains,
and we have added organics and troilite to
the mostly amorphous silicate grains of Uchida et al. (2004).
The optically thin dust temperatures are calculated from radiative
equilibrium at each radius. 
%NC
The results of this simple modeling procedure are shown in Figure 3, 
where we show the spectrum of each individual component;
model parameters are given in Table 1.
It is clear, in the case of {\lk},
the two components alone cannot
account for the 
excess emission above photospheric
fluxes in the near IR; some excess can be
seen in the case of {\gm} and {\dm} as well.
We find that 
this excess can be explained by
black body emission at $T_d = $ 1400K, the
dust destruction temperature
(see Figure 3). We suggest that the innermost
region of the inner disk is optically thick
and has a rim at the dust destruction radius
from where this emission arises, as
is the case in other young disks (Dullemond, Dominik, \& Natta  2001; Muzerolle et al. 2003, M03).
%Assuming that it is blackbody emission, we
%can estimate the the solid angle
%required to fit the near-IR excess.

Several points can be extracted from
our analysis: (1) all disks have
cleared their inner regions up to few AU.\footnote{The location of the
outer edge of the gap $R_t$ for {\gm}
is larger than that inferred by
Rice et al. (2003). These authors only
had broad band colors, and included the
silicate emission as emission from the wall.
With our better observations, the optically
thin emission can be separated from the
wall optically thick emission, which is
much lower in the $\sim 10 \mu$m region.}
%NC 
(2) The height of the wall $z_w$ at $R_t$
puts an upper limit to the height
of the outer disk at that radius because the wall could
be ``puffed up'' with the enhanced radiative
heating. Comparing with
predictions from
disk structure calculations,
the low values of $z_w/R_t$
in {\gm} and {\dm}
are consistent with  models with very significant
grain growth (cf. Figure 5 in D'Alessio et al. 2001); the much lower value
of {\lk} cannot be explained by models with well mixed 
gas and dust; it probably requires a significant amount
of dust settling.
(3) The column density of dust required
to produce the silicate feature is $\Sigma_d \sim \tau_{min} / \kappa(10 \mu{\rm m})
\sim 0.001 \, {\rm gr \, cm^{-2}}$,
with $\kappa(10 \mu{\rm m}) \sim 1 \, {\rm cm^2/gr}$.
Using standard expressions for accretion
disks (cf. M03), we can
estimate the column
density 
of gas in the inner disk from the mass accretion rate $\mdot$.
With $\mdot \sim 3 \times 10^{-9} \msunyr$
as representative of our objects (Table 1),
we obtain
$\Sigma_{gas} \, \sim \, 20 \, {\rm gr \, cm^{-2}}$ at 1 AU, for 
$\alpha = 0.01$, $T =$ 100K, and $M = 1 \msun$,
with corresponding dust column of $\sim 0.2 \, {\rm gr \, cm^{-2}}$
(using the standard dust-to-gas mass ratio),
much higher than detected.
As discussed
in C02, one possibility is that
the remaining dust is locked up in larger bodies with
low near-IR opacities. 
(4) We can estimate the height $z_{dust}$ of the
rim at the dust destruction radius from
the solid angle required to fit the emission,
assuming a cylindrical geometry. The radius
of this cylinder would be the dust destruction
radius, which
can be calculated from the stellar and accretion
luminosities (M03, Table 1).
We find $z_{dust}  \le 0.1 H$ (Table 1),
with $H$ $\sim 0.1 R$, which is
much lower than the height of the rim previously found
in thick inner disks (Dullemond et al. 2001; M03).
We note that black body emission alone can
explain the near-IR excess (Figure 3.) This
suggests that the innermost optically
thick region has a small radial extent and 
moreover, may be in the shadow of the rim (Dullemond et al. 2001).
Taken together, all these results
suggest that the solids in the inner disks of the targets 
have experienced significant evolution.

\begin{deluxetable}{lccc}
\tablewidth{3in}
\tablecaption{Stellar Properties and
Model results}
\tablehead{
\multicolumn{4}{c}{Stellar Properties}\\\hline
\colhead{}&\colhead{GM Aur} &\colhead{LkCa 15} &\colhead{DM Tau}
}
\startdata
SpT &  K3 & K5  & M1 \\
$T_{eff}$ &  4730 & 4350  & 3720 \\
$A_{V}$ & 0.14  & 0.62  & 0.9 \\
$L$ &  0.83 & 0.74  & 0.25 \\
$R_*$ &  1.35 & 1.51  & 1.20 \\
log L$_{acc}$  &-1.149 & -1.055 & -1.749 \\
log ${\mdot}$ &  -8.563 & -8.096 & -9.103 \\
G$_0$ & 340 & 1500 & 240 \\\hline\hline
%\enddata
%\tablehead{
%\colhead{Model results}&\colhead{} &\colhead{} &\colhead{}
%}
%\startdata
\multicolumn{4}{c}{Model Results}\\\hline
$R_t$/AU &  6.5 & 3  & 4 \\
$T_w$/K &  124 & 177  & 117 \\
$z_w/R$ &  0.12 & 0.03  & 0.13 \\
$\tau_{min}$ &  0.005 & 0.007  & 0.007 \\
$R_d/R_*$ &  12 & 9.8  & 12 \\
$z_{dust}/R_d$ &  0.01 & 0.08  & 0.006 \\
\enddata
\tablecomments{G$_0$ = Field at 100 AU in Habing, 1 Habing = standard 
UV flux from interstellar radiation field.
Stellar data in solar units from Kenyon \&
Hartmann (1995).
}
\end{deluxetable}

\section{Discussion}

In a small sample of stars selected solely on the basis of strong
and diverse disk molecular emission we have found two striking results
(1) the presence of strong UV {\em continuum} emission from H$_2$ molecules
and (2) evidence for grain growth and possible inner disk clearing
by planetary bodies.
%in the same systems with the strongest
%H$_2$ emission (\tw , \dm , \gm , and \lk ).   
Although our
sample is limited we suggest that  these two results may be related.  
The process of grain growth and
inner disk clearing will certainly enhance the penetration power of both
UV and X-ray radiation, perhaps even leading to regions that 
are optically thin to UV radiation with penetration only weakly limited by 
molecular opacity.  
The direct or improved  exposure of 
molecular gas to high energy radiation can certainly produce the 
H$_2$ UV emission features seen in our study,
%NC
which only require a small molecular column 
($< 10^{19}$ cm$^{-2}$; \citet{herczeg_twhya2}).
% and only a small 
%of molecular column ($< 10^{19}$ cm$^{-2}$; \citet{herczeg_twhya2}) 
%is required to produce this radiation.   
Additional FUV spectra of young 
TTS combined with forthcoming Spitzer/IRS and BASS data on the mid-IR SED will
be required to fully determine the relation between FUV emission and grain growth. 

Electron impact excitation
requires the mixture of hot electrons within an undissociated H$_2$
layer  (Raymond, Blair, \& Long  1997).   
For these systems we can assume that the
upper layers of the inner disks are mostly cleared of small grains and H$_2$ self-shielding 
limits photo-destruction.
Under normal circumstances radiation will erode
the H$_2$ layer, but these systems are still accreting, which  provides
a source term.  In the shielded layer some  hot photo-electrons will be produced via
UV ionization of C$^{+}$, but these do not have sufficient energy ($\sim 2$ eV) 
for H$_2$ excitation.  Thus, electron excitation likely requires X-ray photons which
have greater penetration power and produce higher energy electrons. 
In the dust poor inner disk the opacity for abundant 1 keV X-rays will also be 
reduced by nearly a factor of 3 \citep{glassgold_xray}
allowing X-rays to penetrate into the self-shielded
layer, ionize H$_2$, and produce hot electrons mixed within molecular gas. 
The radiation produced by electron impact excitation of H$_2$, 
therefore does not originate from the star 
(with an oblique angle of incidence on the
disk like Ly $\alpha$), but rather arises from the upper layers of the disk
itself.  This increases the penetration power of the UV radiation
with resulting effects on  disk physics (UV clearing of inner disk, accretion,
gas temperature structure, etc.)
and the chemistry of species sensitive to radiation below 1700~\AA .

The reduction of dust opacity inside the planet forming regions of the inner disk 
along with the presence of gas (inferred from accretion) suggests that some of the heating 
mechanisms believed important for the outer disk will not be 
effective in the inner disk.  In particular, the efficiency of photoelectric
heating can be expected to be reduced and a greater role must be played by X-ray heating
and UV excited H$_2$ collisional de-excitation.  Thus the 
gas physics of an evolving inner disk will change as a function of dust evolution. 
UV emission from H$_2$ must trace these changes and provides a direct
tracer of the gas conditions within the planet-forming regions of 
T Tauri disks that are only weakly probed by the dust.
%and our results
%provide the first hints of a links between dust evolution, disk viscous
%evolution, and the molecular evolution.

\acknowledgments

We are grateful for constructive and useful comments from an anonymous referee.
EAB and NC are grateful for several discussions with A. Dalgarno.
This work has been supported by
NASA through grant 09374.01-A from the STSCI
and Origins of Solar Systems grant NAG5-9670.
PD acknowledges grants from DGAPA and CONACyT.
For this work MLS was supported in part by NASA grant NAG5-9475 and the 
University Research Council of the Univ. of Cincinnati. DKL and RWR were 
supported by The Aerospace Corporation's Independent Research and Development 
program and by the USAF Space and Missile Systems Center 
through the Mission Oriented Investigation and Experimentation program,
under contract F4701-00-C-0009.

%\clearpage
%\begin{deluxetable}{lccc}
%\tablewidth{0pt}
%\tablecaption{Stellar Properties and Model results}
%\tablehead{
%\multicolumn{4}{c}{Stellar Properties}\\\hline
%\colhead{}&\colhead{GM Aur} &\colhead{LkCa 15} &\colhead{DM Tau}
%}
%\startdata
%SpT &  K3 & K5  & M1 \\
%$T_{eff}/K$ &  4730 & 4350  & 3720 \\
%$A_{V}$ & 0.14  & 0.62  & 0.0 \\
%$L/\lsun$ &  0.83 & 0.74  & 0.25 \\
%$R_*/\rsun$ &  1.35 & 1.51  & 1.20 \\
%log L$_{acc}/\lsun$  &-1.149 & -1.055 & -1.749 \\
%log M$_{\odot}$/M$_{\odot} \, {\rm yr}^{-1}$ &  -8.563 & -8.096 & -9.103 \\
%G$_0$(100AU)/Habings & 340 & 1500 & 240 \\\hline\hline
%%\enddata
%%\tablehead{
%%\colhead{Model results}&\colhead{} &\colhead{} &\colhead{}
%%}
%%\startdata
%\multicolumn{4}{c}{Model Results}\\\hline
%$R_t$/AU &  6.5 & 3  & 4 \\
%$T_w$/K &  124 & 177  & 117 \\
%$z_w/R$ &  0.12 & 0.03  & 0.13 \\
%$\tau_{min}$ &  0.005 & 0.007  & 0.007 \\
%$R_d/R_*$ &  12 & 9.8  & 12 \\
%$z_{dust}/R_d$ &  0.01 & 0.08  & 0.006 \\
%\enddata
%\tablecomments{Stellar data are taken from Kenyon \&
%Hartmann (1995).  1 Habing = standard UV flux from interstellar radiation field.
%}
%\end{deluxetable}

%\clearpage

%Figuras

\end{document}